\documentclass[a4paper,12pt,oneside]{article}
\usepackage[english]{babel}
\usepackage{graphicx}

\begin{document}

\title{ The analysis of magnetic field measurements of T Tau
}

\author{ 
D.A.Smirnov$^1$, M.M.Romanova$^2,$ S.A.Lamzin$^1$
} 

\date{ \it \small
1) Sternberg Astronomical Institute, Moscow V-234, 119992 Russia
lamzin@sai.msu.ru
\\
2) Department of Astronomy, Cornell University, Ithaca, USA
\\
}

\maketitle

\bigskip


\section*{Abstract}

\large{
It is shown that the existence of hot accretion spots at the surface of T
Tau practically has no effect on the accuracy of estimation of its
magnetic field strength at photospheric level.

  We also found that one can interpret results of T Tau's photospheric
magnetic field strength measurements carried out via different methods in
the frame of the following alternative: 1) if T Tau's inclination angle
$i\le 10^\circ,$ then magnetic field of the star may be dipolar with the
angle between rotational and magnetic axes is near
$85^\circ;$ 2) if it will be found (e.g. from interferometric observations)
that $i>10^\circ,$ then magnetic field of T Tau is essentially non-dipolar
or/and non-stationary.
}

\section*{ Introduction }

\large{

  T Tauri stars are low mass $(M\le 2\,M_\odot),$ young $(t<10$ Myr)
pre-main sequence stars. T Tauri itself belongs to subclass of classical
TTS, what means that its activity is the result of (protoplanetary) disc
accretion onto central young star. The problem of strength and geometry of
TTS's magnetic field is one of the most important in the physics of young
stars because it is widely belived that magnetic field determines the
character of TTS's activity and controls evolution of stellar angular
momentum. At the moment the strength of magnetic field is known for less
than ten TTS due to difficulties of magnetic field measurements in these
objects.

  Basing on the fact that magnetic field changes equivalent widths of
spectral lines depending on the value of their Lande factor and line
splitting pattern, Guenther et al. (1999) and Johns-Krull et al. (1999)
found that avereged surface magnetic field strength of T Tau $B> 2$ kG. On
the other hand, Smirnov et al. (2003, 2004) found from specrtopolarimetric
observations that averaged {\it longitudinal} (i.e. line of site oriented)
component of T Tau's surface magnetic field $B_{\|}$ was $\simeq +150\pm 50$
G in 1996 and 2000 yr, and $\simeq +50\pm 30$ G in 2003 yr, i.e. more than
order of magnitude less than $B$-value. Possible reasons of so strong
discrepancy is considered below.


\section*{
The influence of accretion onto accuracy of estimation of
observed $B$ and $B_\|$ values
}

  Calculations of Romanova et al. (2003) demonstrated that when disk
accretion occurs onto a star with dipolar magnetic field gas predominantly
falls near stellar magnetic poles, i.e. onto regions where magnetic field
strength is maximal. Accreted matter heats stellar surface, so if
contribution of "hot"\, spots into formation of some spectral line is
negligible, then average $B$ and $B_\|$ values, derived from observations of
this line, will be less than in the case of star without hot spots. Let
estimate this effect in a quantitave way.

  Effective temperature $T_{ef}$ of the star in the accretion zone can be
estimated from the energy conservation equation:
$$
\sigma T_*^4 + \frac{\rho V^3}{2} \simeq \sigma T^4_{ef}, 
\eqno(1)
$$
where $T_*$ is an effective temperature of the star outside the accretion 
zone; $\rho$ and $V$ are density and infall velocity of accreted matter
respectively. Spectra of regions with $T>T_*$ differ from spectrum of
undisturbed photosphere. To take this effect into account we will suppose in
the following that spectral lines of interest do not originate in these
regions at all.

  We adopted $\rho=\rho (\theta, \varphi)$ and $V=V(\theta,\varphi)$
distributions of infall gas parameters along stellar surface -- $\theta$ and
$\varphi$ are latitude and longitude respectively -- from Romanova et al.
(2003) calculations and used them to estimate $T_{ef}(\theta, \varphi)$
distribution from Eq.(1), assuming $T_*=5250$ K (White \& Ghez, 2001). Three
dimensional magnetohydrodynamic calculations of Romanova et al. (2003) were
carried out for different values of an angle $\alpha$ between magnetic and
rotational axes.
\footnote{
In addition, Romanova et al. assumed that polar field strenght of the star
$B_0$ is $\simeq 2.1$ kG and accretion rate $\dot M_{ac} \simeq 2\cdot 
10^{-7}$ M$_\odot$/yr.
}
Temperature distribution shown in Fig.1 was calculated with $\alpha =
75^{\circ}.$ More precisely, lines of $\Delta T(\theta,\varphi) = T_{ef}
- T_* = const,$ that characterize excess heating of stellar surface in the
accretion zone are shown in this figure. Similar distributions for other
values of $\alpha$-parameter can be constructed from data presented in
Romanova et al. (2004) paper.

  $B_{||}$ and $B$ values that one derives from observation of 
photospheric lines results from averaging of respective local values over
visible hemisphere of the star with radius $R_*$:
$$
B = {1\over 2\pi R_*^2}\, \int_{S_*} B(\theta, \varphi)\, 
\cos \beta \, {\rm d}S, \qquad
B_{||} = {1\over 2\pi R_*^2}\, \int_{S_*} B(\theta, \varphi)\, 
\cos \beta \cos \gamma \, {\rm d}S.
\eqno(2)
$$
Consider spherical system of coordinate with the origin in the
center of the star, $\theta=0$ axis of which coinsides with stellar
rotational axis, characterized by unity $\vec \omega$ vector. Let
$\varphi=0$ plane of this system be the plane that passes throught vector
$\vec \omega$ and unity vector ${\bf l}$ oriented to the Earth. Let also
${\bf n}$ be a unity vector normal to stellar surface at given point and
${\bf B}$ be a vector of magnetic field at the same point. Then in Eq.(2)
$R_*^{-2}\,{\rm d}S = \sin \theta \,{\rm d}\theta \,{\rm d}\varphi;$ 
$\beta$ is the angle between ${\bf n}$ and ${\bf l}$ vectors;
$\gamma$ is the angle between ${\bf l}$ and ${\bf B},$ such as:
$$
\cos \beta = \sin i\, \sin \theta\, \cos \varphi + \cos i \, \cos \theta,
\qquad
\cos \gamma = {\bf l} \cdot { {\bf B} \over B },
$$
where $i$ is the angle between stellar rotational axis and the line of site,
i.e. between $\vec \omega$ and ${\bf l}$ vectors.

  ${\bf B}(\theta, \varphi)$ vector will vary with time $t$ due to stellar
rotation. Let assume that magnetic field of the star is dipolar, i.e.
$$
{ {\bf B} \over B_0 } = 
{ 3 {\bf n} (\vec \mu\,{\bf n}) - {\vec \mu} \over 2 }, 
$$
where $B_0$ is {\it polar} magnetic field strength and ${\vec \mu}$ is the
unity vector oriented along stellar magnetic axis. Then to discribe ${\bf
B}(t)$ dependence in each point of stellar surface one needs only one
parameter -- phase $\psi$ $( 0\le \psi <1 ),$ that characterise rotation of
stellar magnetic pole relative to $\varphi=0$ plane. Accretion flow disturbs
initial (dipolar) magnetic field of the star, but it follows from
calculations of Romanova et al., (2003) that the difference between initial
and disturbed fields near stellar surface does not exceed 1\,\%.

  To estimate the effect of hot spots existence on the average values of
total magnetic field strength and its longitudinal component we calculated
"observed"\, values $B_{||}$ and $B$ by means of Eq.(2) for three cases: 1)
neglecting with existence of hot spots, i.e. calculating integrals over all
visible hemisphere of the star; 2) excluding regions of stellar surface
(accretion zone) where $\Delta T>200$ K; 3) excluding regions where $\Delta
T>1000$ K;

  Variations of $B_{||}/B_0$ value with phase of stellar rotational period
$\psi$ for these three cases are shown in Fig.2 as an example of our
calculations carried out with $\alpha = 75^{\circ}$ and four values of
inclination angle $i:$ $10^{\circ}$, $30^{\circ}$, $60^{\circ}$ and
$90^{\circ}.$ Analogous variations $B(\psi)/B_0$ for the same values of
$\alpha$ and $i$ parameters are shown in Fig.3. It can be seen from these
figures that the difference between curves does not exceed $10-15\,\%$ at
any value of $i.$ Similar result was found at other values of $\alpha$
parameter. It means that errors in measured values of $B$ and $B_{||}$ due
to neglecting of existence of hot spots is less than $15\,\%,$ what is less
than modern accuracy of measurements of these values in TTSs. It is the
consequence of relatively small $(\le 10$\,\%) area of stellar surface
occupied by accretion zone, where the main portion of accretion energy is
liberated (Romanova et al. 2003; 2004).


\section*{ Estimation of parameters of T Tau's magnetic field under
assumption that it is dipolar}

   Let assume that global magnetic field of T Tau is dipolar and stationary.
Then more than order of magnitude difference between observed values of
$B_{\|}$ and $B$ is due to proximity of angle $\delta$ between magnetic
axis and line of sight to $90^\circ$ at the moment of observations. (At
$\delta=90^\circ$ longitudinal field components that parallel and
antiparallel to the line of sighte are mutially compensated and average
value of $B_{\|}$ should be exactly equal to zero.) Consider this question
in a quantitative way basing on the results of existing measurement.

   To discribe dipolar magnetic field one needs three parameters: $B_0,$
$i$ and $\alpha.$ For given set of these parameters observed (average)
values of $B$ and $B_\|$ will vary with phase $\psi,$ e.g. as shown in
Fig.2 and Fig.3. Let suppose that probability $p$ to detect value $B$ in the
range $B^{obs}\pm 3\,\sigma$ at some moment is equal to the lenght of phase
interval, inside which $\mid B - B^{obs} \mid<3\, \sigma,$ where $\sigma$ is
the error of measurements. Probability to detect value $B_\|$ in the range
$\mid B_\| - B_\|^{obs} \mid<3\,\sigma$ is determined in a similar way.

  Guenther et al. (1999) and Johns-Krull et al. (1999) found $B=2.3\pm 0.15$
kG and the following values of $B_{\|}$ was found in our observations:
$+150\pm 50$ Гс in 1996 yr and 2002 yr as well as $+50\pm 30$ G in 2003 yr.
If $B_0$ and $\alpha$ parameters did not vary with time (magnetic field is
stationary) then the difference in observed values of $B$ and $B_\|$
is the consequence of observations at different phases $\psi.$ Then one can
find for any set of three parameters $B_0,$ $i$ and $\alpha$ probability
$P=p_1\cdot p_2\cdot p_3\cdot p_4\cdot p_5$ to detect two $B$-values and
three $B_\|$ that was indeed found from observations.

  Such calculations were carried out in the following range of parameters:
$B_0$ from 1.5 to 5 kG, $0^\circ \le \alpha\le 90^\circ$ and $0^\circ \le 
i \le 60^\circ.$ We took into account that $i={29^o\,}^{+10^o}_{-15^o}$
according to Akeson et al. (2002) when adopting the range for $i.$
Calculations indicated that:

  1) Probability $P$ differs from zero only if 3\,kG\,$< B_0 <$ 4.3\, kG at
any values of $\alpha$ and $i$ from the range of interest. This result looks
natural if to take into account that surface magnetic field varies between
$B_0$ (at magnetic pole) and $B_0/2$ (at magnetic equator) in the case of
dipol, so average B-value should be inside the range $(B_0/2,B_0).$

  2) $P=P(\alpha,i)$ function is practically independent on values of
$B_0$ from allowable range. As an axample we plot in Fig.4 $P=P(\alpha,i)$
function calculated at $B_0=3.5$ kG. (Lines of equal probability consist of
segmnets of stright lines because $P(\alpha,i)$ was calculated at discrete
values of $\alpha$ and $i$ angles with $5^\circ$ step.)

  As follows from Fig.4 at $\alpha < 80^\circ$ and/or $i>10^\circ$
probability to find observed values of $B$ and $B_\|$ is less than 0.2, so
one can consider this possibility as a low probability one. (Note by the way
that this conclusion is practically independent on the real value of T Tau's 
$B_0.$) Two conclusions follow from this statement:

  1) If future observations (e.g. with VLTI interferometer) will demonstrate
that inclination angle of T Tau is larger than $10^\circ,$ then existing
results of $B$ and $B_\|$ measurements mean that magnetic field of
the star is essentially non-dipolar near its surface. Note that there are
some theoretical reasons to belive that configuration of magnetic fields of
TTSs can be close to quadrupol (Dudorov, 1995).

2) If it will be found that $i<10^\circ,$ then existing measurements of
magnetic field in T Tau do not contradict to the hypotesis that it has
dipolar magnet field with dipol axis inclined to the rotational axis at
$\alpha \simeq 85^\circ.$

  Note once more that these statements are valid only if magnetic field of 
T Tau is stationary, otherwise there is no sence to compare values of
$B$ and $B_\|$ that was observed non-simultaneously.


\section*{ Conclusion }

  Calculations of Romanova et al. (2003) indicate that in the case of disc
accretion onto inclined magnetic dipol, with parameters $B_0$ and $\dot
M_{ac}$ typical to classical TTS, the difference between resulting and
undisturbed magnetic field near stellar surface is negligible. On the other
hand we found that in the case of dipolar magnetic field the existence of
hot accretion spots at stellar surface practically has no effect on the
accuracy of measurements of average $B$ and $B_\|$ values.

  It means that comparison of observed $B$ and $B_\|$ values at photospheric
level could give information to understand if TTS's magnetic field is
dipolar or has more complicate geometry. We demonstrated above that in the
case of T Tau one can agree observed values of $B$ and $B_\|$ in the frame
of the following alternative: magnetic field of the star is eccentially
non-dipolar or it is dipolar, but dipol axis is inclined to stellar
rotational axis at $\simeq 85^\circ.$ It will be possible to discriminate
between these possibilities when future (interferometric) observations will
answer the question: does inclination $i$ of stellar rotation axis to the line
of sight exceeds $10^\circ$ or not.

  It would be more precisely to say that such alternative follows from
observations only if magnetic field of T Tau is stationary. If it does
indeed -- this question is interesting by itself but to answer it more
observations are necessary.

\bigskip

{\it
  This work was supported by grants RFBR 02-02-16070, INTAS 03-51-6311, NASA
NAG5-13220, NAG5-13060, NSF AST-0307817.
}

\newpage

\begin{figure}[h!]
 \begin{center}
  \resizebox{12.5cm}{!}{\includegraphics[angle=90]{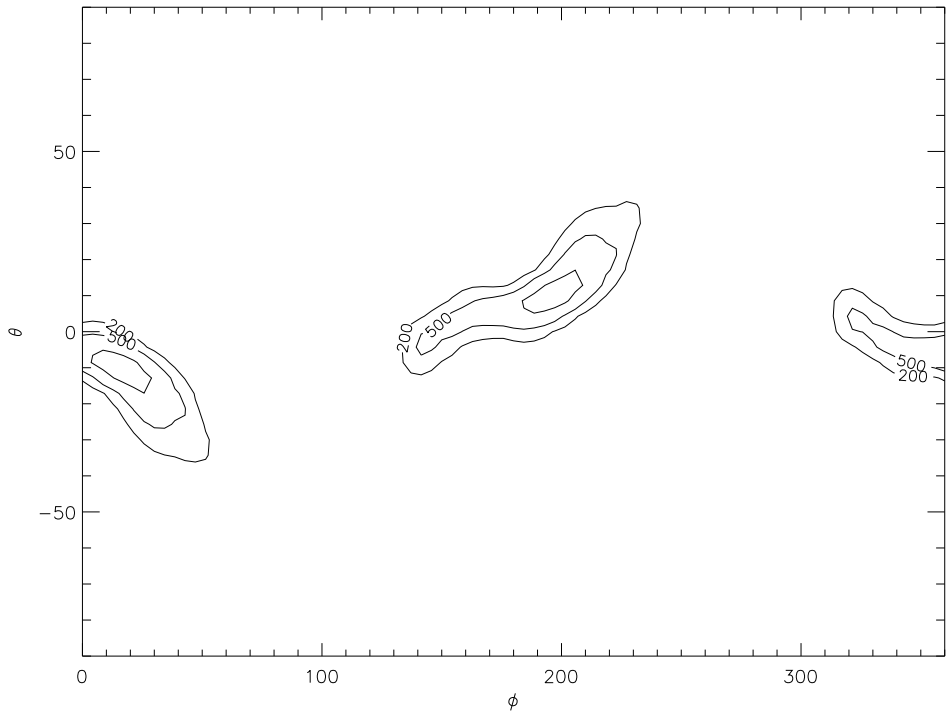}}
  \caption{
Distribution of excess temperature $\Delta T$ on the surface of a star that
has dipolar magnetic field with the angle between magnetic and rotational
axes $\alpha = 75^{\circ}$ (see text for details). The innermost countour
respects to $\Delta T=1000$ K. Longitude and latitude (in degrees) are
plotted along X and Y axes respectively.
            }
  \end{center}
\end{figure}


\newpage

\begin{figure}[t!]
 \begin{center}
  \resizebox{14.5cm}{!}{\includegraphics[angle=0]{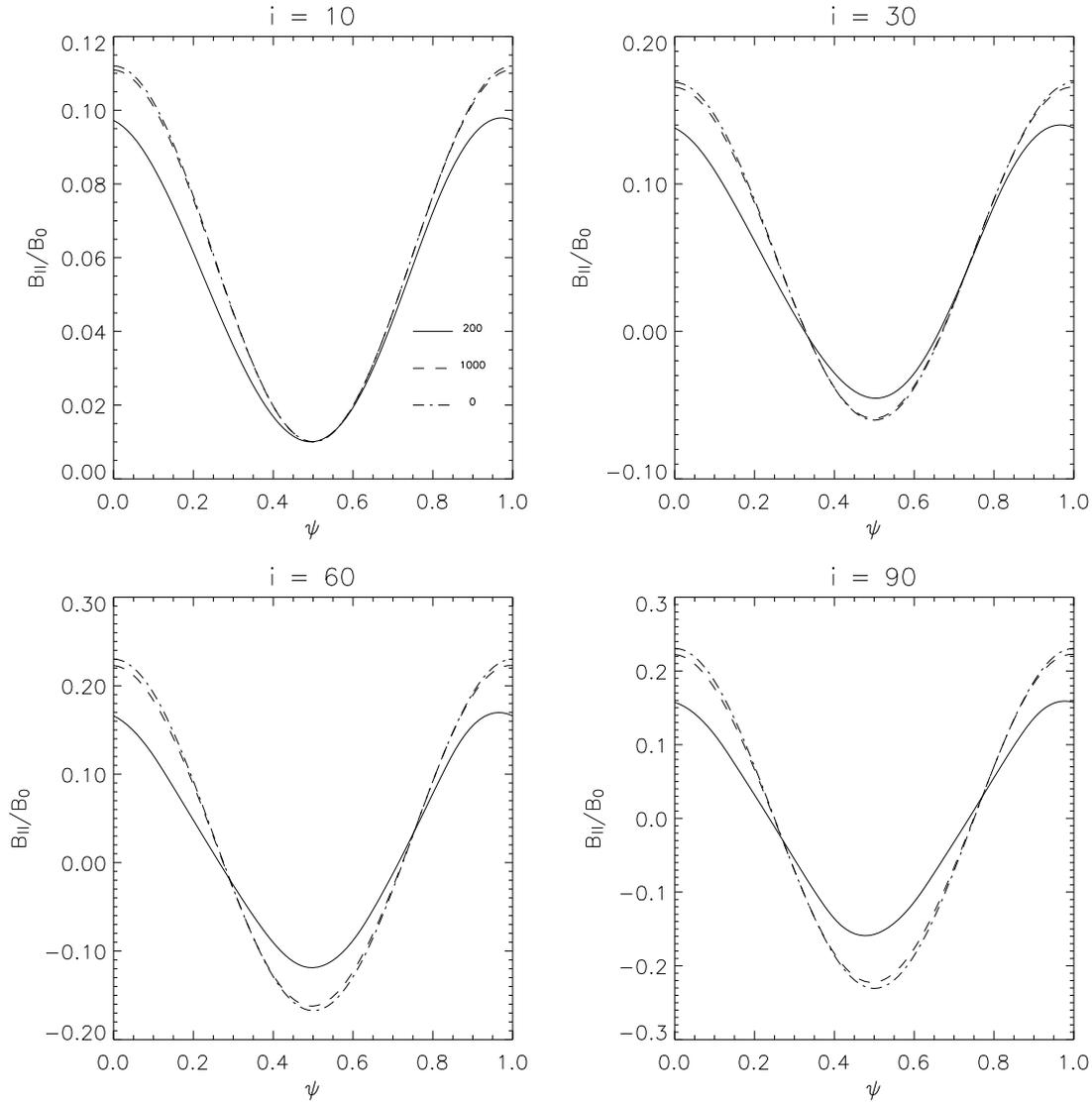}}
  \caption{
Longitudinal component $B_{\|}$ of stellar magnetic field strength as
function of phase of rotational period $\psi$ for $\alpha=75^\circ$ and
different values of inclination angle $i.$ Dot-dashed lines respects to
situation when all stellar surface is considered (there is no hot spots),
solid lines -- stellar surface without regions with $\Delta T> 200$ K,
dashed lines -- stellar surface without regions with $\Delta T> 1000$ K.
            }
  \end{center}
\end{figure}

\newpage

\begin{figure}[t!]
 \begin{center}
  \resizebox{14.5cm}{!}{\includegraphics[angle=0]{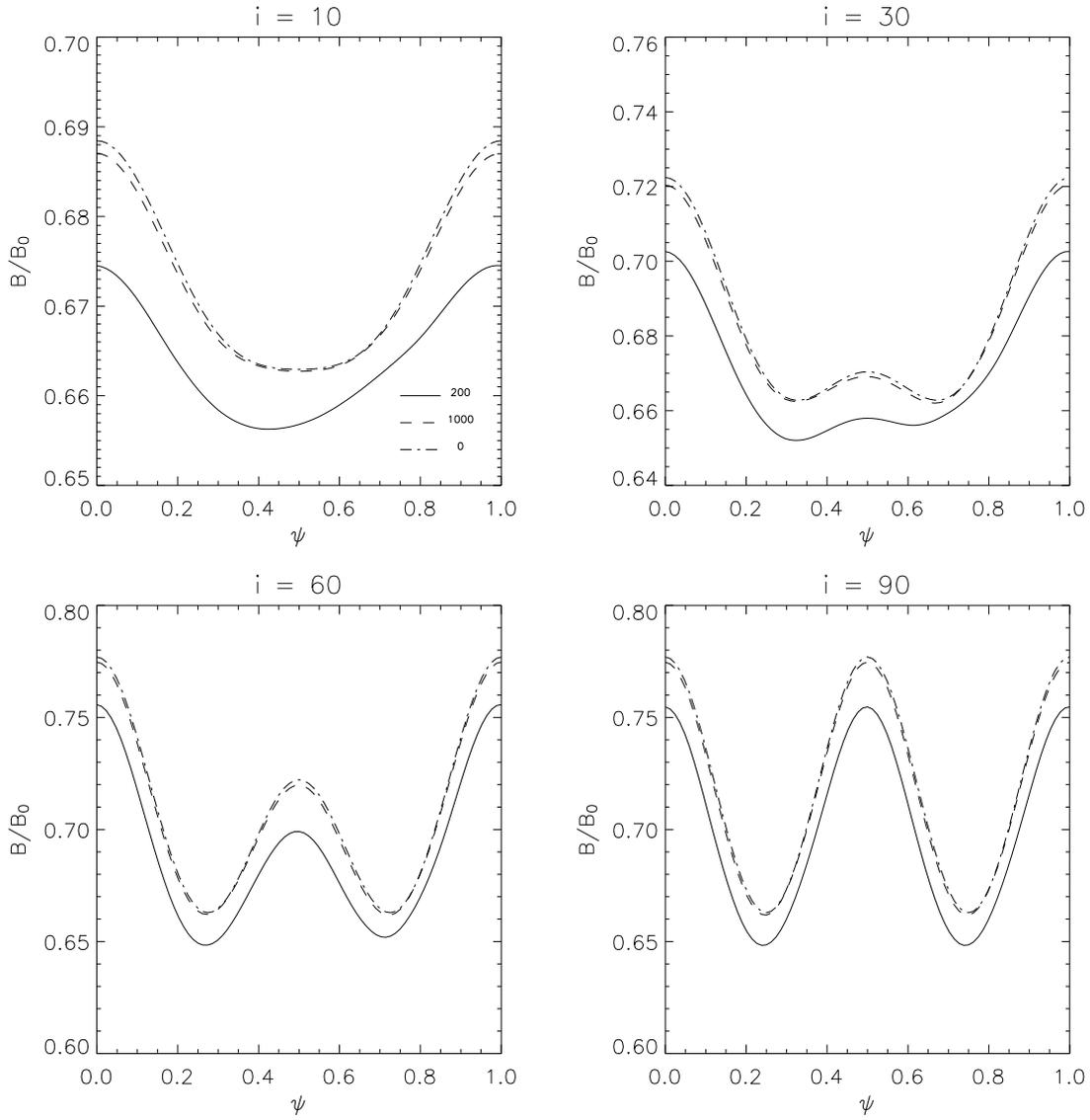}}
  \caption{The same as in Fig.2 but for average magnetic field strength $B.$
            }
  \end{center}
\end{figure}
\newpage
\begin{figure}[t!]
 \begin{center}
  \resizebox{8.5cm}{!}{\includegraphics[angle=90]{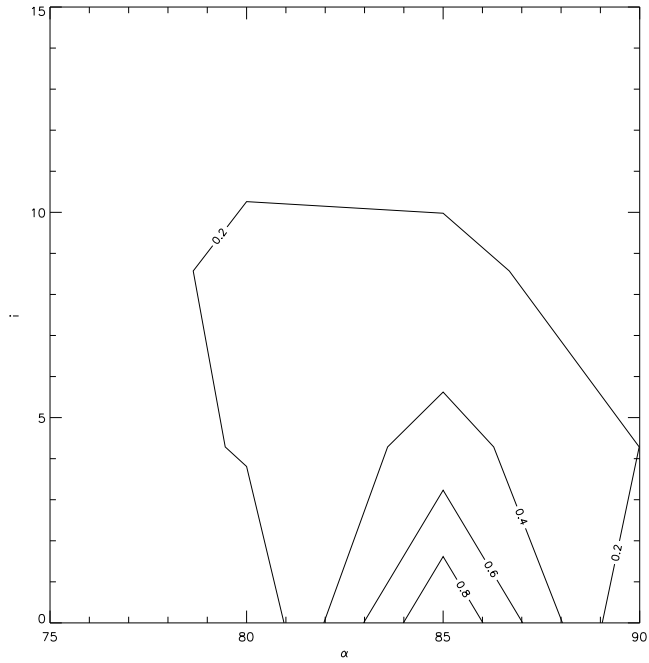}}
  \caption{
Lines of equal probability $P(\alpha,i)=const$ in the case when $B_0=3.5$
kG. See text for details.
            }
  \end{center}
\end{figure}

\end{document}